\documentstyle [preprint,tighten,aps,epsf] {revtex}
\begin {document}

\title {Addition spectra of quantum dots: the role of dielectric mismatch}
\author {Alberto Franceschetti, Andrew Williamson and Alex Zunger}
\address {National Renewable Energy Laboratory, Golden, CO 80401}
\maketitle

\begin {abstract}
Using atomistic pseudopotential wave functions
we calculate the electron and hole charging energies
of InAs quantum dots. We find that the charging energies
depend strongly on the dielectric constant $\epsilon_{\rm out}$ of the
surrounding material, and that when the latter is smaller
than the dielectric constant of the dot (weak external
screening) the electron-electron and hole-hole interactions
are dominated by surface polarization effects.
We predict the addition energies and the quasi-particle gap
as a function of size and $\epsilon_{\rm out}$.
We find excellent agreement with recent single-dot
tunneling spectroscopy data for $\epsilon_{\rm out} = 6$.
\end {abstract}

\newpage

\vskip 0.5cm

Semiconductor quantum dots can be made with various dielectric coatings:
Organic molecules \cite {Murray,Guzelian},
other semiconductors (e.g. self-assembled dots \cite {SK}, 
core-shell nanocrystals \cite {Coreshell}, 
lithographically-etched dots \cite {Tarucha},
strain-induced dots \cite {Strain}), or
glasses \cite {Glass}.
It has been realized \cite {Keldish,Brus} that the dielectric environment 
can profoundly affect the optical and transport 
properties of quantum dots.
This can be seen by considering the two processes described in Fig. 1,
where a quantum dot of dielectric constant $\epsilon_{\rm in}$ is embedded
in a material of dielectric constant $\epsilon_{\rm out}$.
Figure 1(a) depicts the process of adding three electrons to
an otherwise neutral quantum dot.
The initial configuration of the system, of energy $E_0$, 
consists of a neutral dot in the ground state
and a Fermi reservoir at the reference energy $\varepsilon_{\rm ref} = 0 $.
The ``charging energy'' $\mu_1$ required to load the 
first electron into the quantum dot is

\begin {equation} \label {Mu1}
\mu_1 \equiv E_1 - E_0 = \varepsilon_{e1} + \Sigma_{e1}^{\rm pol},
\end {equation} 

\noindent where $E_1$ is the total energy of the dot with one additional electron,
$\varepsilon_{e1}$ is the energy of the single-particle level $e1$
with respect to the reference energy $\varepsilon_{\rm ref}$,
and $\Sigma_{e1}^{\rm pol}$ is the self-energy of the additional electron
interacting with its own image charge created by
the dielectric mismatch at the surface of the dot \cite {Brus}.
The charging energy $\mu_2$ to add the second electron to the quantum dot is

\begin {equation} \label {Mu2}
\mu_2 \equiv E_2 - E_1 = \varepsilon_{e1} + \Sigma_{e1}^{\rm pol} + J_{e1,e1} \, ,
\end {equation}

\noindent where $J_{e1,e1}$ is the Coulomb interaction between the two electrons.
It includes a direct electron-electron contribution $J^{\rm Coul}_{e1,e1}$
and a polarization contribution $J_{e1,e1}^{\rm pol}$
arising from the interaction of one
electron with the image charge of the other electron \cite {Brus}.
Finally, the charging energy for the third electron is

\begin {equation}
\mu_3 \equiv E_3 - E_2 = \varepsilon_{e2} + \Sigma^{\rm pol}_{e2} + 2J_{e1,e2} - K_{e1,e2} \, ,
\end {equation} 

\noindent where $K_{e1,e2}$ is the exchange energy between two electrons
with parallel spins in the $e1$ and $e2$ single-particle levels.
The ``addition energies'' for the second and the third electrons are, respectively

\begin {eqnarray} 
\Delta_{1,2} & \equiv & \mu_2 - \mu_1 = J_{e1,e1} \, , \label {D12} \\
\Delta_{2,3} & \equiv & \mu_3 - \mu_2 = 
(\varepsilon_{e2} - \varepsilon_{e1}) + 
(\Sigma^{\rm pol}_{e2} - \Sigma^{\rm pol}_{e1}) + \nonumber \\
& & (2J_{e1,e2} - J_{e1,e1}) - K_{e1,e2} \, . \label {D23}
\end {eqnarray}

\noindent Since $\Sigma^{\rm pol}_{i}$ and $J^{\rm pol}_{i,j}$ depend
strongly on the dielectric constant of the surrounding material,
the charging spectroscopy \cite {Tarucha,Banin}
of a quantum dot depends on its dielectric environment.

Figure 1(b) describes the process of removing an electron from
the highest occupied orbital of a neutral quantum dot 
and placing it  into the lowest unoccupied orbital of an identical dot
(located at infinite distance from the first dot).
The energy required by this process (``quasi-particle gap'') is the difference between the ionization
potential and the electron affinity of the dot.
The initial configuration, consisting of the two neutral dots in the ground state,
has energy $2E_0$, while the final configuration has energy $E_1 + E_{-1}$, 
where $E_{-1}$ is the energy of the quantum dot with a hole 
in the highest occupied orbital $h1$.
The quasi-particle gap is then 

\begin {equation} \label {Eqp}
\varepsilon_{\rm gap}^{\rm qp} = E_{1} + E_{-1} - 2 \, E_{0} =
(\varepsilon_{e1} - \varepsilon_{h1}) + \Sigma_{e1}^{\rm pol} + \Sigma_{h1}^{\rm pol} \, ,
\end {equation}

\noindent where $\varepsilon_{\rm gap} \equiv \varepsilon_{e1} - \varepsilon_{h1}$
is the single-particle (HOMO-LUMO) gap.
We see that the quasi-particle gap depends, via the polarization
self-energies $\Sigma_{e1}^{\rm pol}$ and $\Sigma_{h1}^{\rm pol}$,
on the dielectric environment.
The optical gap differs from the quasi-particle gap by the electron-hole interaction
$J_{h1,e1}$:

\begin {equation} \label {Eopt}
\varepsilon_{\rm gap}^{\rm opt} = \varepsilon_{\rm gap}^{\rm qp} - 
(J^{\rm Coul}_{h1,e1} + J^{\rm pol}_{h1,e1}).
\end {equation}

\noindent Very recently single-dot tunneling spectroscopy
was applied to InAs nanocrystals \cite {Banin}, measuring
$\Delta_{N,N+1}$ and $\varepsilon_{\rm gap}^{\rm qp}$. 

The effects of dielectric confinement on the excitonic gap and the charging energies
of quantum dots have been addressed in the past
\cite {Brus,Babic,Banyai,Takagahara,Lannoo,Goldoni}
using the effective-mass approximation (EMA).
Recent pseudopotential calculations \cite {Exciton}
have demonstrated the importance
of using an atomistic description of the quantum dot electronic structure
for calculating the electron-hole Coulomb energy.
The pseudopotential approach provides an accurate
description of the wave function decay outside
the quantum dot and of the interband coupling due to quantum confinement,
which are critical for a correct evaluation
of the polarization and Coulomb energies in small nanocrystals. 

Using pseudopotential wave functions,
we discuss here the effects of dielectric mismatch ($\epsilon_{\rm out} \ne \epsilon_{\rm in}$) on
(a) the electron and hole charging energies $\mu_{N} = E_{N} - E_{N-1}$,
where $E_N$ is the ground-state energy of $N$ electrons (or holes)
in the quantum dot,
(b) the addition energies $\Delta_{N,N+1} = \mu_{N+1} - \mu_{N}$, and
(c) the quasi-particle band gap $\varepsilon_{\rm gap}^{\rm qp}$.
We find excellent agreement with recent experimental results \cite {Banin},
and interpret the data in terms of Coulomb and polarization  contributions 
[Eqs. (\ref {Mu1}) - (\ref {Eqp})].
We show that depending on the ratio
between $\epsilon_{\rm in}$ and $\epsilon_{\rm out}$ 
one encounters two physically
distinct regimes of transport behavior:

(i) When $\epsilon_{\rm out} \ll \epsilon_{\rm in}$ (weak external screening)
the electron-electron interaction $J_{i,j}$ is dominated by
the polarization contribution $J_{i,j}^{\rm pol}$, which depends only weakly on the single-particle
states $i$ and $j$.
The charging energies $\mu_N$ depend strongly
on the dielectric constant $\epsilon_{\rm out}$, and 
are widely spaced for various $N$ (large addition energies $\Delta_{N,N+1}$). 

(ii) When $\epsilon_{\rm out} \ge \epsilon_{\rm in}$ (strong external screening)
the dominant contribution to $J_{i,j}$ is the Coulomb energy $J_{i,j}^{\rm Coul}$,
which is quite sensitive to the identity of the states $i$ and $j$.
The charging energies $\mu_N$ depend weakly
on $\epsilon_{out}$ and are more closely spaced (small addition energies).

The practical significance of these results stems from the fact
that, due to the long-range character of the Coulomb interaction
and the exponential decay of the wave functions outside
the quantum dot, 
dielectric confinement and quantum confinement can be physically separated.
In fact, by changing the dielectric environment {\it far away}
from the dot, while keeping the same barrier material next to the dot,
one can control and tailor 
the electronic properties (such as $\Delta_{N,N+1}$ and $\varepsilon_{\rm gap}^{\rm qp}$)
without affecting quantum confinement (i.e. the single-particle energies and wave functions).

We approximate the many-particle wave function $\Psi_N$ of a system of $N$ electrons
in the conduction band of a quantum dot
by a single Slater determinant constructed from the
wave functions $\{\psi_i, i=1 \cdot \cdot \cdot N \}$ 
of the $N$ single-particle states occupied by the $N$ electrons.
The corresponding total energy is

\begin {equation} \label{Eel}
E_{N} = E_{0} + \sum _{i} (\varepsilon_i + \Sigma_i^{\rm pol}) \, n_i +
\sum_{i < j} (J_{i,j} - K_{i,j}) \, n_i \, n_j \; ,
\end {equation}

\noindent where $\varepsilon_{i}$ are the conduction-band 
single-particle energy levels,
$\Sigma_i^{\rm pol}$ are the polarization self-energies,
$J_{i,j}$, $K_{i,j}$ are the electron-electron
Coulomb and exchange energies, respectively,
and $n_i$ are the occupation numbers ($\sum_i \, n_i = N$).
The ground state $\Psi_N^0$ corresponds to the configuration 
that minimizes the total energy $E_N$. 
In Eq. (\ref {Eel}) we neglect:
(i) the coupling between different Slater determinants 
(i.e. configuration-interaction effects), and
(ii) the response of the single-particle wave functions
$\psi_i$ to the electrostatic field
(i.e. self-consistent effects).
These assumptions are sufficiently accurate in small, three-dimensional
structures in the strong-confinement regime \cite {Exciton,CI,Rontani}. 

The single-particle energies 
$\varepsilon_i$ and wave functions $\psi_i ({\bf r}, \sigma)$
are obtained here from the solution of the Schroedinger equation:

\begin {equation} \label {Schr}
[ -\nabla^2 + V_{ps} ({\bf r}) ] \, \psi_i ({\bf r}, \sigma) =
\varepsilon_i \, \psi_i ({\bf r}, \sigma) .
\end {equation}

\noindent The pseudopotential of the quantum dot $V_{\rm ps} ({\bf r})$ is obtained from
the superposition of screened atomic potentials,
which are fitted \cite {Williamson} to reproduce the bulk experimental
optical transition energies and effective masses, as well
as the surface work function.
Spin-orbit coupling is fully included in the solution of the Schroedinger equation.

The interelectronic energies $J_{i,j}$ are given by:

\begin {equation} \label {Jij}
J_{i,j} = e \, \sum_{\sigma} \int |\psi_i ({\bf r}, \sigma)|^2 \,
\Phi_j ({\bf r}) \, d{\bf r} ,
\end {equation}

\noindent where $\Phi_j ({\bf r})$ is the electrostatic potential
energy due to a charge distribution 
$\rho_j ({\bf r}) = e \, \sum_{\sigma} |\psi_j ({\bf r}, \sigma)|^2$
in a dielectrically inhomogeneous medium.
$\Phi_j ({\bf r})$ satisfies the Poisson equation:

\begin {equation} \label {Poisson}
\nabla \cdot  \epsilon({\bf r}) \nabla \Phi_j ({\bf r}) =
- 4 \pi e \, \rho_j ({\bf r}) ,
\end {equation}

\noindent where $\epsilon ({\bf r})$ is the (position-dependent)
{macroscopic} dielectric constant of the system. 
The Poisson equation is solved on a real-space
grid using a finite-difference discretization of the gradient operator.
The boundary conditions are obtained from a multipole expansion of the electrostatic potential \cite {Exciton}.
The dielectric constant $\epsilon({\bf r})$ changes smoothly from 
$\epsilon_{\rm in}$ to $\epsilon_{\rm out}$,
with a transition region of the order of the interatomic bond-length.
The interelectronic energy $J_{i,j}$ can be separated into two contributions:
(a) the direct Coulomb energy $J_{i,j}^{\rm Coul}$, which corresponds to the interaction
between two electrons in the quantum dot {as if} the dielectric constant was
uniform throughout the system, and identical to the macroscopic dielectric constant
of the quantum dot; 
and (b) the polarization energy $J_{i,j}^{\rm pol}$ which
accounts for the effects of the dielectric discontinuity
at the interface between the quantum dot and the surrounding material,
and the ensuing surface polarization charge.

The polarization self-energies $\Sigma_i^{\rm pol}$ are given by:

\begin {equation} \label {Sigma}
\Sigma_i^{\rm pol} = {e \over 2} \sum \nolimits _{\sigma} \int 
\psi_i^* ({\bf r}, \sigma) \, \Sigma({\bf r}) \, \psi_i ({\bf r}, \sigma)
\, d{\bf r} ,
\end {equation} 

\begin {equation}
\Sigma({\bf r}) = \lim_{{\bf r}' \rightarrow {\bf r}}
[ G({\bf r}, {\bf r}') - G_{\rm bulk} ({\bf r}, {\bf r}')] ,
\end {equation}

\noindent where $G({\bf r}, {\bf r}')$ is the Green's function associated
with the Poisson equation [Eq. (\ref {Poisson})], and 
$G_{\rm bulk} ({\bf r}, {\bf r}')$ is the bulk Green's function.
Here we use the analytical expression of $\Sigma ({\bf r})$ 
for a spherical quantum dot \cite {Banyai} of dielectric constant $\epsilon_{\rm in}$
embedded in a medium of dielectric constant $\epsilon_{\rm out}$.
The singularity of $\Sigma ({\bf r})$ at the surface of the dot is removed
by applying a smoothing function $1 - e^{-(r-R)^2/\sigma^2}$, where $\sigma$ 
is a broadening factor of the order of the bond length.

We consider InAs spherical nanocrystals of diameter
$D = 30.3$ and $42.2 \, {\rm \AA}$.
The surface dangling bonds are passivated using a large-gap barrier material \cite {Williamson}.
Our analysis of the envelope functions
extracted for the pseudopotential wave functions shows
that the first electron level ($e1$) is predominantly $s$-like,
while the next 3 electron levels ($e2$, $e3$, and $e4$) are
predominantly $p$-like. The first two hole levels ($h1$ and $h2$)
have an $s$-like envelope function, while the next two hole levels
($h3$ and $h4$) have a $p$-like envelope function.
Each single-particle energy level is doubly degenerate
(because of time-reversal symmetry). 

The self-energies $\Sigma_i^{\rm pol}$, the 
polarization energies $J_{i,j}^{\rm pol}$, and the
Coulomb energies $J_{i,j}^{\rm Coul}$
of the $30.3 \, {\rm \AA}$ diameter
InAs nanocrystal are shown in Fig. 2 as a function of the
external dielectric constant $\epsilon_{\rm out}$, for a few 
single-particle states $i$ and $j$. 
We see that (i) both $\Sigma_i^{\rm pol}$ and
$J_{i,j}^{\rm pol}$ depend strongly on $\epsilon_{\rm out}$, and vanish
when $\epsilon_{\rm out} = \epsilon_{\rm in}$ (vertical arrows in Fig. 2).
(ii) When $\epsilon_{\rm out} > \epsilon_{\rm in}$
the polarization energies $J_{i,j}^{\rm pol}$ become negative, thus acting to diminish 
the electron-electron interaction.
(iii) The dependence of $\Sigma_i^{\rm pol}$ and $J_{i,j}^{\rm pol}$
on the identity of the orbitals $i$ and $j$ (e.g. $s$ or $p$) 
is rather weak, as shown in the insets in Fig. 2.
(iv) There is a critical value of $\epsilon_{\rm out}$ 
($\epsilon_{\rm critical} \sim 4$) such that for 
$\epsilon_{\rm out} < \epsilon_{\rm critical}$
the polarization energies $J^{\rm pol}_{i,j}$ dominate over 
the Coulomb energies $J^{\rm Coul}_{i,j}$.

%The conductance spectrum of a quantum dot as a function of the gate voltage $V_g$
%displays peaks \cite {Tarucha,Banin} in correspondence to the charging energies:
%$V_{g,N} = \lambda \, \mu_N$, where $\lambda$ is a scale factor
%that depends on the distance between the gate and the dot. 
The charging energies $\mu_N = E_N - E_{N-1}$, calculated 
from the total energies $E_N$ given by Eq. (\ref {Eel}),
are shown in the central panel of Fig. 3 as a function of $\epsilon_{\rm out}$.
The vertical arrow at the bottom of the figure
denotes the value $\epsilon_{\rm out} = \epsilon_{\rm in}$,
which divides the behavior into two domains:
(i) In the weak screening regime ($\epsilon_{\rm out} \ll \epsilon_{\rm in}$)
the charging energies are widely
spaced, and their value depend strongly on $\epsilon_{\rm out}$.
(ii) In the strong screening regime, on the other hand, the charging
energies are closely spaced, and do not depend significantly
on $\epsilon_{\rm out}$. The calculated charging spectrum is shown in Fig. 3
for $\epsilon_{\rm out} = 1 $ (left-hand side) and 
$\epsilon_{\rm out} = 20$ (right-hand side), illustrating these two
limiting behaviors.

The electron and hole addition energies $\Delta_{N,N+1}$ 
(spacings between peaks in the charging spectra of Fig. 3)
are given in Table I for a few values
of $\epsilon_{out}$. We see that 
(i) for a given value of $\epsilon_{\rm out}$,
the addition energy for the third electron $\Delta_{2,3}^{(\rm e)}$
is significantly larger than the addition energy for the second electron $\Delta_{1,2}^{(\rm e)}$.
This can be explained by noting from Eqs. (\ref {D12}) and (\ref{D23})
that while $\Delta_{1,2}^{(\rm e)}$ measures only the interelectronic repulsion,
$\Delta_{2,3}^{(\rm e)}$ includes also the single-particle gap 
between the $s$-like state $e1$ and the $p$-like states $e2$, $e3$, and $e4$
($\varepsilon_{e2} - \varepsilon_{e1} = 0.40 \, {\rm eV}$ for the $D=30.3 \, {\rm \AA}$ nanocrystal
and $0.36 \, {\rm eV}$ for the $D=42.2 \, {\rm \AA}$ nanocrystal).
(ii) The addition energies of the remaining electrons (up to $N = 8$)
are approximately constant. The addition energy of the 9-th electron $\Delta_{8,9}^{(\rm e)}$ 
is slightly larger, and reflects the single-particle gap between the $p$-like shell
and the next shell.
(iii) While Coulomb blockade effects were often interpreted in terms of the ``constant interaction'' model,
whereby the addition energies $\Delta_{N,N+1}$ are constant and independent of $N$,
we see that in small nanocrystals deviations from this model are noticeable, 
and are due primarily to the existence of single-particle gaps
comparable or larger than the interelectronic energies $J_{i,j}$.
These deviations are magnified when polarization effects are small 
($\epsilon_{\rm out} \ge \epsilon_{\rm in})$.
(iv) The electron and hole addition energies become smaller as the size $D$ increases and/or the dielectric
constant $\epsilon_{\rm out}$ decreases.
(v) The quasi-particle gap $\varepsilon_{\rm gap}^{\rm qp}$
depends strongly on $\epsilon_{\rm out}$, while the optical gap $\varepsilon_{\rm gap}^{\rm opt}$ does not.
This is so because in the optical gap 
$\Sigma_{h1}^{\rm pol} + \Sigma_{e1}^{\rm pol}$ and  $J_{h1,e1}^{\rm pol}$
tend to cancel [see Eq. (\ref {Eopt})], so
$\varepsilon_{\rm gap}^{\rm opt} \sim (\varepsilon_{e1} - \varepsilon_{h1}) - J_{h1,e1}^{\rm Coul}$.
(vi) The optical gap is smaller than the quasi-particle gap. For $\epsilon_{\rm out} \sim \epsilon_{\rm in}$
the difference is $J_{h1,e1}^{\rm Coul}$.

Figure 3 and Table I provide clear predictions of the charging energies
and addition energies of InAs quantum dots.
To compare with the experimental measurements of Banin {\it et al.} (Ref. \onlinecite {Banin}),
in which $\epsilon_{\rm out}$ is an unknown quantity, we first fit our calculated
$\Delta_{1,2}^{(\rm e)}$ for the smaller dot with the experimental value 
$\Delta_{1,2}^{(\rm e)} = 0.22 \, {\rm eV}$, 
finding that $\epsilon_{\rm out} = 6$ gives a good fit (Table I).
Using this value of $\epsilon_{\rm out}$, we then predict for $D = 30.3 \, {\rm \AA}$
(experimental data in parenthesis for $D = 34 \, {\rm \AA}$) 
$\varepsilon_{\rm gap}^{\rm qp} = 1.78$ $(1.75)$,
$\Delta_{1,2}^{(\rm h)} = 0.24$ $(0.20)$,
$\Delta_{2,3}^{(\rm h)} = 0.23$ $(0.22)$,
$\Delta_{2,3}^{(\rm e)} = 0.64$ $(0.71)$,
and $\Delta_{3,4}^{(\rm e)} = 0.24$ $(0.23)$.
Using the same value of $\epsilon_{\rm out}$,
our predictions for $D = 42.2 \, {\rm \AA}$
(experimental data in parenthesis for $D = 44 \, {\rm \AA}$) are:
$\varepsilon_{\rm gap}^{\rm qp} = 1.38$ $(1.38)$,
$\Delta_{1,2}^{(\rm h)} = 0.16$ $(0.20)$,
$\Delta_{2,3}^{(\rm h)} = 0.16$ $(0.17)$,
$\Delta_{1,2}^{(\rm e)} = 0.15$ $(0.14)$,
$\Delta_{2,3}^{(\rm e)} = 0.51$ $(0.52)$,
and $\Delta_{3,4}^{(\rm e)} = 0.14$ $(0.14)$.
We see that we have excellent agreement with experiment. Our theory
can be further used to decompose the experimentally measured quantities
into distinct physical contributions.
For example, for $D = 30.3 \, {\rm \AA}$ 
the quasi-particle gap $\varepsilon_{\rm gap}^{\rm qp} = 1.78 \, {\rm eV}$ 
includes [Eq. (\ref {Eqp})]
the single-particle gap $\varepsilon_{e1} - \varepsilon_{h1} = 1.71 \, {\rm eV}$
and the self-energy contribution
$\Sigma_{h1}^{\rm pol} + \Sigma_{e1}^{\rm pol} = 0.07 \, {\rm eV}$.
The addition energy for the third electron 
$\Delta_{2,3}^{(\rm e)} = 0.64 \, {\rm eV}$ includes 
[Eq. (\ref {D23})]
the single-particle contribution $\varepsilon_{e2} - \varepsilon_{e1} = 0.40 \, {\rm eV}$,
the Coulomb contribution $2J_{e1,e2}^{\rm Coul} - J_{e1,e1}^{\rm Coul} = 0.17 \, {\rm eV}$,
the polarization contribution $2J_{e1,e2}^{\rm pol} - J_{e1,e1}^{\rm pol} = 0.07 \, {\rm eV}$,
and a negligible self-energy contribution $\Sigma_{e2}^{\rm pol} - \Sigma_{e1}^{\rm pol}$.
The exchange contribution $K_{e1,e2}$ is smaller than  $0.02 \, {\rm eV}$, and can be neglected.  

In conclusion, we predict the effects of the dielectric environment
on the electron and hole charging energies and on the addition spectrum of semiconductor
quantum dots. We find that the charging energies and the addition energies
depend sensitively on the dielectric constant $\epsilon_{\rm out}$
of the surrounding material via the self-energies $\Sigma_i^{\rm pol}$ 
and the polarization energies $J_{i,j}^{\rm pol}$.
When $\epsilon_{\rm out} \ll \epsilon_{\rm in}$ the charging energies
are widely spaced in energy, and depend strongly on $\epsilon_{\rm out}$.
When $\epsilon_{\rm out} \ge \epsilon_{\rm in}$ the charging energies
are more closely spaced. Our calculations are in  excellent
agreement with recent spectroscopic results \cite {Banin}
for $\epsilon_{\rm out} = 6$.

This work was supported by the U.S. DOE, OER-BES, Division
of Materials Science, under Grant No. DE-AC36-98-GO10337.

%%%%%%%%%%%%%%%%%%%%%%%%%%%%%%%%%%%%%%%%%%%  TABLES %%%%%%%%%%%%%%%%%%%%%%%%%%%%%%%%%%%%%%%%%%%%%%%%%%%%%%%%%%%%

\begin {table}
\caption {Addition energies $\Delta_{N,N+1}$, quasi-particle gap $\varepsilon_{\rm gap}^{\rm qp}$,
and optical gap $\varepsilon_{\rm gap}^{\rm opt}$ of InAs nanocrystals (in eV) as a function
of the dielectric constant $\epsilon_{\rm out}$.}
\begin {tabular} {ccccccccc}
& \multicolumn {4} {c} {D = $30.3 \, {\rm \AA}$} & \multicolumn {4} {c} {D = $42.2 \, {\rm \AA}$} \\
& \multicolumn {4} {c} {$\varepsilon_{gap} = 1.71 \, {\rm eV}$} & \multicolumn {4} {c} {$\varepsilon_{gap} = 1.31 \, {\rm eV}$} \\
\noalign {\vskip 0.1cm}
\tableline
\noalign {\vskip 0.1cm}
$\epsilon_{\rm out}$ & $1$ & $6$ & $10$ & $20$ & $1$ & $6$ & $10$ & $20$ \\
\noalign {\vskip 0.1cm}
\tableline
\noalign {\vskip 0.3cm}
\multicolumn {3} {l} {\bf Electrons} \\
  \noalign {\vskip 0.1cm}
  $\Delta_{1,2}^{(\rm e)}$          & 0.96 & 0.22 & 0.16 & 0.11  & 0.69 & 0.15 & 0.10 & 0.07 \\ 
  $\Delta_{2,3}^{(\rm e)}$          & 1.45 & 0.64 & 0.57 & 0.53  & 1.05 & 0.51 & 0.46 & 0.43 \\ 
  $\Delta_{3,4}^{(\rm e)}$          & 0.99 & 0.24 & 0.18 & 0.13  & 0.69 & 0.14 & 0.10 & 0.07 \\
  $\Delta_{4,5}^{(\rm e)}$          & 0.98 & 0.23 & 0.17 & 0.12  & 0.70 & 0.15 & 0.10 & 0.07 \\ 
  $\Delta_{5,6}^{(\rm e)}$          & 0.99 & 0.24 & 0.18 & 0.13  & 0.69 & 0.15 & 0.10 & 0.07 \\ 
  $\Delta_{6,7}^{(\rm e)}$          & 0.99 & 0.24 & 0.18 & 0.13  & 0.69 & 0.14 & 0.10 & 0.06 \\
  $\Delta_{7,8}^{(\rm e)}$          & 0.99 & 0.24 & 0.18 & 0.13  & 0.69 & 0.15 & 0.10 & 0.07 \\ 
  $\Delta_{8,9}^{(\rm e)}$          & 1.04 & 0.28 & 0.21 & 0.17  &&&& \\
  $\Delta_{9,10}^{(\rm e)}$         & 1.00 & 0.25 & 0.19 & 0.14  &&&& \\
\noalign {\vskip 0.3cm}
\multicolumn {3} {l} {\bf Holes} \\
  \noalign {\vskip 0.1cm}
  $\Delta_{1,2}^{(\rm h)}$          & 0.98 & 0.24 & 0.18 & 0.13  & 0.71 & 0.16 & 0.12 & 0.09 \\ 
  $\Delta_{2,3}^{(\rm h)}$          & 0.97 & 0.23 & 0.17 & 0.12  & 0.71 & 0.16 & 0.12 & 0.08 \\ 
  $\Delta_{3,4}^{(\rm h)}$          & 0.98 & 0.24 & 0.18 & 0.13  & 0.71 & 0.16 & 0.12 & 0.09 \\ 
  $\Delta_{4,5}^{(\rm h)}$          & 1.02 & 0.23 & 0.16 & 0.12  & 0.73 & 0.17 & 0.13 & 0.09 \\ 
  $\Delta_{5,6}^{(\rm h)}$          & 0.97 & 0.23 & 0.16 & 0.12  & 0.71 & 0.16 & 0.12 & 0.08 \\ 
\noalign {\vskip 0.3cm}
\multicolumn {3} {l} {\bf Gaps} \\
  \noalign {\vskip 0.1cm}
  $\varepsilon^{\rm qp}_{\rm gap}$  & 2.37 & 1.78 & 1.72 & 1.65  & 1.84 & 1.38 & 1.32 & 1.27 \\  
  \noalign {\vskip 0.1cm}
  $\varepsilon^{\rm opt}_{\rm gap}$ &      & 1.56 & 1.55 & 1.54  &      & 1.22 & 1.21 & 1.20 \\
\end {tabular}
\end {table}

%%%%%%%%%%%%%%%%%%%%%%%%%%%%%%%%%%%%%%%%%%%  FIGURES %%%%%%%%%%%%%%%%%%%%%%%%%%%%%%%%%%%%%%%%%%%%%%%%%%%%%%%%%%%%

\begin {figure}
\centerline {\epsffile {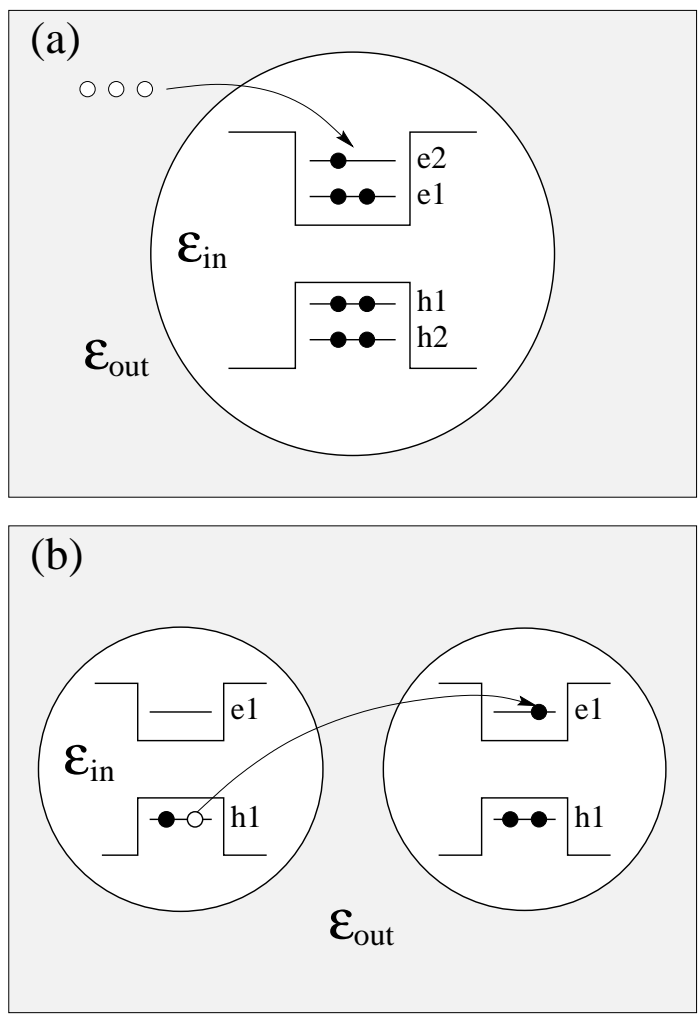}}
\caption {Part (a) illustrates the process of loading three electrons into an otherwise neutral quantum dot.
Part (b) shows the process of removing a single electron
from a dot and placing it into another dot.}
\end {figure}

\begin {figure}
\centerline {\epsffile {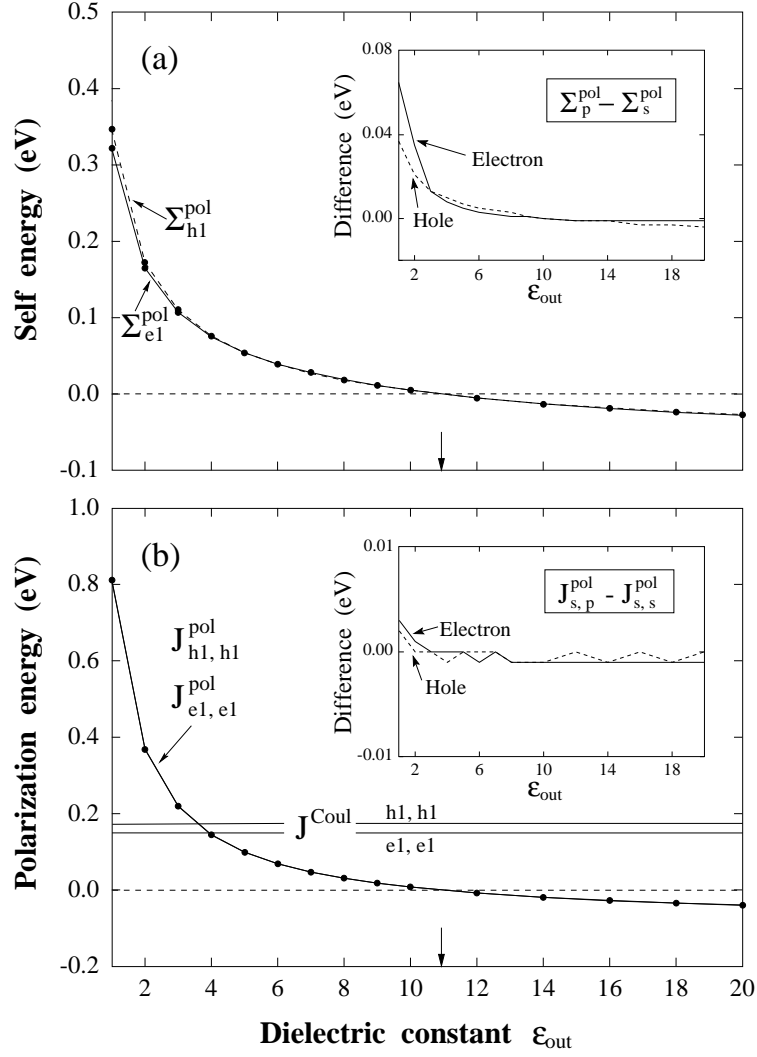}}
\caption {The self-energy $\Sigma_i^{\rm pol}$ and the polarization
energy $J_{i,j}^{\rm pol}$ of an InAs quantum dot
(diameter $D = 30.3 \, {\rm \AA}$) are shown as a function
of the outside dielectric constant $\epsilon_{\rm out}$.
The insets show the differences $\Sigma_p^{\rm pol} - \Sigma_s^{\rm pol}$
and $J_{s,p}^{\rm pol} - J_{s,s}^{\rm pol}$
as a function of $\epsilon_{\rm out}$. 
The vertical arrows indicate the value $\epsilon_{\rm out} = \epsilon_{\rm in}$.}
\end {figure}

\begin {figure}
\centerline {\epsffile {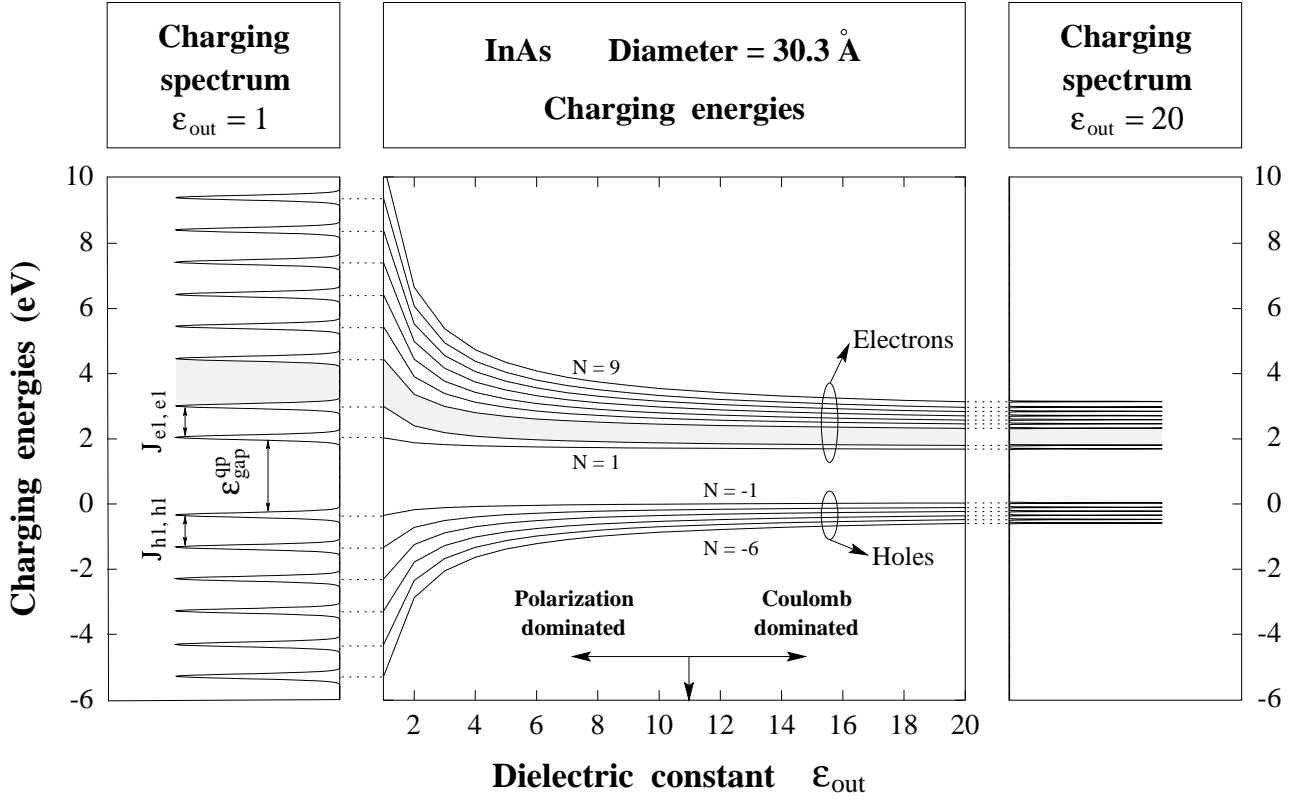}}
\caption {The central panel shows the dependence of a the electron charging energies
$\mu_1 \cdots \mu_9$ and of the hole charging energies $\mu_{-1} \cdots \mu_{-6}$
on the external dielectric constant $\epsilon_{\rm out}$.
The vertical arrow indicates the value $\epsilon_{\rm out} = \epsilon_{\rm in}$.
The side panels show the calculated charging spectrum in the case $\epsilon_{\rm out} = 1$
(left-hand panel) and $\epsilon_{\rm out} = 20$ (right-hand panel). 
The zero of the energy scale corresponds to the highest energy valence state.}
\end {figure}

\end {document}